\def\ba{\bar a}

\def\tt{\tilde t}

\def\be{\begin{equation}}
\def\ee{\end{equation}}
\def\te{\end{equation}}
\def\bea{\begin{eqnarray}}
\def\ba{\begin{eqnarray}}
\def\eea{\end{eqnarray}}
\def\ea{\end{eqnarray}}
\def\tea{\end{eqnarray}}

\def\1/2{\frac{1}{2}}

\def\ba{\bar a}

\def\tt{\tilde t}

\def\be{\begin{equation}}
\def\ee{\end{equation}}
\def\te{\end{equation}}
\def\bea{\begin{eqnarray}}
\def\ba{\begin{eqnarray}}
\def\eea{\end{eqnarray}}
\def\ea{\end{eqnarray}}
\def\tea{\end{eqnarray}}

\def\(#1){(\ref{#1})}

\newskip\humongous \humongous=0pt plus 1000pt minus 1000pt

\newif\ifdtup

\documentclass[a4paper]{jpconf}

\makeatletter                    
\@addtoreset{equation}{section}  
\makeatother                     

\begin{document}

\title{Emergence: Key physical issues for deeper philosophical inquiries}
\author{B. L. Hu}
\address{Maryland Center for Fundamental Physics and Joint Quantum Institute,\\
University of Maryland, College Park, Maryland 20742, USA and \\
Institute for Advanced Study and Department of Physics, \\
Hong Kong University of Science and Technology, Hong Kong, China. }
\ead{blhu@umd.edu}

\begin{abstract}
 A sketch of  three senses of emergence and a suggestive view on the emergence of time and the direction of time is presented. After trying to identify which issues philosophers interested in emergent phenomena in physics view as important I make several observations pertaining to the concepts, methodology and mechanisms required to understand emergence and describe a platform for its investigation.
I then identify some key physical issues which I feel need be better appreciated by the philosophers in this pursuit.
I end with some comments on one of these issues, that of coarse-graining and persistent structures.
\end{abstract}


\centerline{\small {\sl Invited Talk at the Heinz von Foerster Centenary International Conference on Self-Organization and Emergence: }}
\centerline{\small {\sl Emergent Quantum Mechanics (EmerQuM11). Nov. 10-13, 2011, Vienna, Austria. Proc. in J. Phys. (Conf. Series)}}

\section{Introduction}
In this talk I present some latest thoughts on three inter-related subjects:

{\small 1) \textbf{Emergence}: After describing three \textit{different senses of emergence},
I point out that effective field theory (EFT) or renormalization
group (RG) is a suitable, maybe even necessary,  but not sufficient set of conceptual
means for describing emergence. EFT or RG [A1] may suggest
how different physics manifest at different scales, but one also
needs to identify the\textit{ mechanisms or processes} whereby different
levels of structures and the laws governing them, including the
symmetry principles, emerge. That depends on deeper interplay
of collectivity, complexity, stochasticity and self-organization.

2) \textbf{Emergent Gravity}: There are at least two intimately related
veins in viewing gravity as emergent: a) ``General Relativity as
Hydrodynamics?"  (This viewpoint which can be traced back to Sakharov (1968) \cite{Sak} was first pronounced in this vein in \cite{GRhydro}. Other major proponents of this view are Volovik \cite{Volovik} and Wen \cite{Wen}.)--
in the sense that gravity is an effective theory
valid only at the long wavelength, low energy limit of some underlying
theory (quantum gravity \cite{Oriti}) for the microscopic structures of
spacetime and matter [A2][A3].  b) Gravity as Thermodynamics , where
such a view is often shaped by considering the effects of an event
horizon on the quantum fluctuations of a field, shown by Bekenstein \cite{Bek72}, Hawking \cite{Haw75}, Unruh \cite{Unr76} and others, which underlies what is known today as the holography principle \cite{holography}. This view is represented by the works of Jacobson \cite{Jacobson}, Padmanabhan \cite{Pad10} and Verlinde \cite{Ver10} [A4].

3) \textbf{Gravity and Thermodynamics}: Since both gravity and
thermodynamics are classical theories of macroscopic structures,
if a deep connection exists, we should be able to see their direct
relation at this level, without relying on arguments invoking the
microscopic structure of matter (quantum fluctuations). This I first
posed as a challenge in [A5], one which physicists in the 19c in
principle may be able to resolve. If we can meet this challenge we may see the simpler and deeper connection between gravity and thermodynamics without invoking quantum mechanics. If we fail
we will perhaps see more clearly the essential role of quantum
physics in explaining gravity and the necessary implication that a)
either the macroscopic world is fundamentally quantum. (For viewing spacetime as a condensate, see, e.g, \cite{STcond}) or b) quantum mechanics is also emergent from a deeper structure \cite{Adler}, a representation of stochastic processes \cite{Nelson}, or as a form of organizational rules like statistical mechanics \cite{tHooft}.

\vskip .2cm
\noindent -------------

\noindent [A1] For a recent meeting on this topic,
see, e.g.,\\ http://www.perimeterinstitute.ca/Events/
Emergence-and-Effective-Field-Theories/ Schedule/

\noindent [A2]  For references to writings of the major proponents see, e.g., \cite{E/QG,EGrev}.

\noindent [A3] It is easier and more natural for string theorists to view gravity as emergent. See, e.g., \cite{Seiberg,HorPol}

\noindent [A4] For references to earlier work and a critique, see, e.g., \cite{Mario}

\noindent [A5] B. L. Hu, ¡\textit{°Gravity and Thermodynamics: What exactly do we want?}¡± Invited talk at ESF Exploratory Workshop: Gravity and Thermodynamics, SISSA Sept 8, 2011. Further discussion can be found in \cite{DICE12}.}


\vskip .2cm

\centerline{*******}

\vskip .2cm

The above is the Abstract of my talk. In writing up my report, rather than a sketchy summary of all the points raised above, I thought perhaps it is more useful to select one topic and go deeper. Of these, since \textit{emergence} is the overarching issue receiving increasing attention in physics and beyond, in particular, philosophy, I will focus on this subject matter.
In keeping with the selected focus of this paper I will leave my Point 1) in the Abstract as is, pending future elaboration.
Point 2) above has been discussed in recent meetings and the reader can find more details and other authors' related work in the references given in [A2]{A5].

On Point 3) about the { challenge} I posed:  ``Can we deduce a relation between Gravity and Thermodynamics without invoking any quantum consideration?" either one succeeds or one fails to provide such a direct link without relying on quantum arguments, the implications for theoretical physics are equally significant.  If one can meet this challenge , that thermodynamics rules can be used to understand or even derive gravity without appeal to the microphysical constituents of matter or spacetime and the physical laws governing them, it would be a very important step forward. If on the contrary this bridge between gravity and thermodynamics requires quantum mechanics to build, as is assumed from Bekenstein, Hawking onward, then it says something about the role of quantum mechanics in the macroscopic realm, including, and particularly important for spacetimes in this regard.


This paper is organized as follows: In Sec. 2 I give a sketch of the three senses of emergence \footnote{These are thoughts I have toyed with  on and off in the past two decades since the 1991 Huelva Workshop on The Physical Origin of Time-Asymmetry and the 1993 Santa Fe Institute Workshop on `Fluctuations and Forms'.} which await further development.   In Sec. 3 I present some notes on  essays written by philosophers trying to understand  emergence in the physics context for the purpose of identifying what issues these philosophers view as important in their inquires.
In Sec. 4  I make several observations pertaining to the concepts, methodology and mechanisms required to understand emergence and describe a platform for its investigation, namely, nonequilibrium statistical mechanics (e.g., \cite{Zub74,Balescu,Zwa01,CH08}). In Sec. 5 I try to identify some key physical issues which I feel need be better appreciated by the philosophers who wish to deepen their inquiries and bring themselves closer to practising scientists pursuing research topics bearing on this issue. I end with a brief description of the issue of coarse-graining and persistent structures.


\section{Emergence: Processes and Mechanisms}

In this section I jot down some notes  on the different senses of emergence which underlie different  scientific disciplines, as well as a view of emergent time. These are merely sketches of ideas for an idea (as Wheeler liked to say), most likely not new, but which I find useful as nucleus for gathering amorphous threads of thoughts for further analysis.

\subsection{Three Senses of Emergence}
\vskip .2cm

\noindent 1. Emergence in the sense of \textit{difference in manifestations}-- Role of coarse-graining:

* Different manifestations at different levels of structures, hierarchical in form, and corresponding interactions.

* Requires the identification of the range and precision of measurement, thus interfaces necessarily with an observer's probing ability and observation range. Effective field theory has this concern.

* Stability of emergent structures depends on the degree of repeated specific coarse-graining.

* Robustness of emergent structures against the variation of different coarse-graining measures.

(Vaguely, in philosopher's language,  the first two points bring out the issues of \textit{novelty}, and the latter two touch on \textit{autonomy}. More in the next two sections.)

This aspect of emergence is sensitive to the conditions for the appearance of meta-stable structures. The key issues are embodied in studies of the emergence of quasi-classical domains \cite{GelHar93} in environment-induced decoherence \cite{ZurPT,QdecBook} and decoherent  or consistent histories \cite{OmnesBook,GriffBook} formulation of quantum mechanics. Historically these studies focused more on the quantum aspects (transition to classicality), but I want to point out that issues in nonequilibrium statistical mechanics enter in a major way.  See, e.g., Sec. 5 of \cite{timeasy} and \cite{DowKent}.

\vskip .2cm

\noindent 2. Emergence in the sense of \textit{conditioning or control}-- Role of stochasticity:

* Selection by the system's environment through their interactions. For quantum open systems and decoherence studies this would correspond loosely to the functionality of the ``pointer basis" \cite{Zur81}. A particular set of bases is `selected' by the way a system interacts with its environment. It is more sensitive to the \textit{type} of coupling than the \textit{magnitude} of interaction.

* Adaptation of a biological system to its environment is well-known in evolutionary theory. The emergence of new species is the outcome of both genetic (intrinsic) variation and environmental (extrinsic) selection, the former arising from mutation, a stochastic variable \footnote{This is similar to the environment-induced decoherence of a quantum system although the stochastic element there is from the noise and fluctuations in an environment and the quantum to classical transition is on the whole not a stochastic process.}.  Note,  even if a reduced density matrix is almost diagonal with respect to some pointer basis, there is still the question of how its diagonal components, the classical probability or the weighing functions, are distributed, which influences the outcome of emergent entities and behavior.

\vskip .2cm

\noindent 3. Emergence in the sense of \textit{dynamics ( sequencing, `updates', but not time evolution)}:

Here I have in mind models from

* \textbf{Cellula automata} (CA) \cite{Wolfram}: Simple conjunctive rules between neighboring elements when iterated a large number of steps can lead to very different structures -- many will terminate but some will keep evolving.  There is no time, thus no dynamics in terms of time evolution, simply sequencing.  Those which survive can be viewed as emergent structures since they possess both the qualities of being novel and autonomous. But note that in this sense, given the conjunctive rules of the cells, one can predict exactly the outcome of `evolution' (thus not in the biological sense 2 above) without any element of stochasticity involved, even though the outcome may appear irregular or seemingly unpredictable.

* \textbf{Growth and form} in driven diffusive systems \cite{Zia}, self-organization: Different forms emerge arising from the interplay of stochasticity and nonlinearly. Role of nonlinearity in self-organization is widely known.  Role of stochasticity (noise and fluctuations) in the genesis of new forms appear in new branches of nonequilibrium physics- based fields in the 80-90s such as  soft matter physics. I want to add that the effect of memory is also important. (Protein-folding is a well-known example.) This calls for recognizing the importance of non-Markovian processes in emergent phenomena \footnote{For example: every snowflake has its own distinct identity, being a fine record of the density, humidity, temperature as the condensate traverses the different layers of water vapor, growing into a flake.}. Non-Markovian (or memory-laden) processes, or histories (usually connoted as with memories) do not have to refer to dynamics in the sense of time-evolution. It could also be understood in the sense of `sequencing' referred to above in the example of cellula automata.

\subsection{Sense of Time in Emergence}
\vskip .2cm
The essence or relevance of time is in its ordering function. Conceptually, let us consider three distinct elements: a) configurations in each step as basic entities, b)  sequencing or iteration of steps (try to avoid using the word `update', since date is a marking of time).  An example of a): Energetics of phase transition using the free energy density functional $F(T)$ describes critical phenomena, not critical dynamics. It is a one parameter (temperature $T$) family of curves, where the minima signify the existence of meta-stable states. One can discuss the probability of tunneling from one such state to another in this (assuming multi-dimensional space) landscape, but there is no dynamics involved.   Using CA as an example of b), the conjunction rules produce one configuration in one step, then another after the second step, which make up a sequence. There are sequences which terminate after relatively few iterations, others sustain many iterations and  begin to produce various forms. There is still no sense of time in the sense of dynamics yet.

The third ingredient c) is the key. In simple cases like CA, iteration in increasing steps can be viewed as time, as the word `update' conjures. But it is more challenging to understand time in more complex systems where they can interact with various environments. Using biological systems as example, replication continues the species  and provides the stability. This is like the sustaining sequences in CA -- those long-living sequences as measured by the number of iterations. Sufficient stability is necessary for a notion of time to emerge.  Mutations in the genes of a system provides the species a wider sampling pool to adapt to a changing environment. But this stochastic element also introduces a source of instability. Thus even though the sequencing function could facilitate  a notion of micro-time this stochasticity and instability at the cellula level undermines the macro-time as we experience it which requires some sustained unidirectional development.

Note at this stage there is no `arrow of time'  yet. Survival of the species is the precondition for the notion of `direction of time' to enter.  Fluctuation triggers instability in the system in varying degrees: the adverse ones could bring about extinction of the species, the `beneficial' ones enable the species to better adapt and continue to evolve. A sense of `direction of time' or a sense of (forward) history comes only when there is sufficient `success' as measured by the species' ability to survive, thrive and improve on itself. The word `advanced' in advanced civilization connotes a sense of positive direction of evolution. Thus it not only reflects  the passive conditioning by the environment but it also allows for some degrees of regulation, control and adjustment by the system itself.  This amounts to biasing a random process.

Therefore, in summary, {\it sustained sequencing with some degree of stability is the precondition for the notion of time to emerge and the direction of time emerges only with positive feedback in the sustained sequencing of biased random processes}.


The emergence of time and the sense of direction of time are explained above by using the paradigms of evolutionary biology.  Placing these issues  in the physics context, concepts and techniques in nonequilibrium statistical mechanics including open systems dynamics and stochastic processes enter in essential ways. This  will be discussed in a later section. Suffice it here to say that time asymmetry in many physical processes is influenced by many factors:  the way one stipulates the boundary conditions and initial states, the time scale of observation in comparison to  the
dynamical time scale, how one decides what the most suitable relevant variables
are and how they are separated from the irrelevant ones, how the irrelevant variables are coarse-grained, and what assumptions one makes and what limits one takes in shaping a macroscopic picture from one's imperfect knowledge of the underlying microscopic structure and dynamics. More discussions follow in Sec. 4, 5.

\section{Philosophical Studies}
\vskip .2cm

In this section I gather some notes in a naive attempt to extract the wisdom from philosophers  trying to understand  emergence in the physics context. I begin with the rudimentary exercise of finding out what the philosophers interpret the physicists (I mean by necessity a selected subset sampling) in their use of words like  ``constructivism" ``protectorate" and the philosophers' terminology ``microphysicalism" ``supervenience" and key words used by both communities, such as ``novelty" ``autonomy" which characterize emergence. Of course these exercises are not just for clarifying the terminology, but for extracting the meaning behind these words. What I find in this  preliminary exercise is that the philosophers are  mostly interested in issues of their own, in metaphysics, in the philosophy of the mind, etc, not unexpectedly. From the questions they raised I also see some ambiguity (or not clear enough definitions) in the physicists' use of words like ``fundamental", ``reductionism", which need be clarified. The focusses of  philosophers of science with a physics background or those who are attuned to physicists' language or way of thinking enable us to see what issues they consider as important to bring back to their own community. At the same time   I also see the philosophers' choice of focus is somewhat partial, or skewed, missing some key issues which physicists consider as important in understanding emergent phenomena. This is likely a result of the scarcity of explanations by the practising physicists of their technical findings in the light of emergence phenomena, even scarcer in philosophical terms.

\subsection{Sources and Tracks}
\vskip .2cm

I shall first identify a few sources of literature on both sides so readers can see how my readings could be limited or biased and  identify what I have missed out clearly.

Classic papers in physics:  Anderson \cite{And72}, Laughlin and Pines \cite{LP} on the one side and Weinberg  \cite{Weinberg} on the other.

Physicists (at PI workshop [A1]) who addressed key issues on emergence: Goldenfeld, Kadanoff

Issues of emergence exemplified in quantum decoherence: Stamp

Philosophers  (at PI workshop) who raised key issues on emergence: Batterman \cite{Batt}, Morrison \cite{Morr}

Philosophers with physics training or attuned to physical issues;  Bain \cite{BainEFT,BainQHE}, Wilson \cite{Wilson}.

Useful introduction: Mainwood PhD thesis \cite{Main}, with metaphysical emphasis.

\vskip .2cm

\subsection{Terminology, Contents, Meaning and Issues}
\vskip .2cm
Physicists usage:  \textbf{`Fundamental', Reductionism, Constructivism, ``protectorate"} \cite{LP}

Emergent entities, properties, principles (Weinberg)

\noindent Philosopher's usage:  \textbf{microphysicalism, supervenience}

\vskip .2cm

\noindent Defining properties of emergence largely agreed-upon by philosophers:

\noindent \textbf{Novelty, Autonomy}(reductive, predictive, causal and/or explanatory)

A description rather than a definition is probably best to illustrate what novelty refers to:

{\small  \textit{`Instead, at each level of complexity entirely new properties appear' }(Anderson \cite{And72}, p. 393).}

{\small \textit{``When you put enough elementary units together, you get something
that is more than the sum of these units. A substance made of a great
number of molecules, for instance, has properties such as pressure and
temperature that no one molecule possesses. It may be a solid or a
liquid or a gas, although no single molecule is solid or liquid or gas." }(Wheeler and Ford \cite{WheFor} p.341, quoted by Mainwood \cite {Main} p.18)}

\vskip .2cm
\noindent \textbf{Reductionism}-- reduction, prediction; derivable, deducible:

{\small \textit{ ``In most philosophical discussions the concept of emergence is intimately related to the following notions: antireductionism, unpredictability, and novelty. Emergence in these contexts is also typically associated with parts and wholes. The idea being that a phenomenon is emergent if its behavior is
not reducible to some sort of sum of the behaviors of its parts, if its behavior is not predictable given full knowledge of the behaviors of its parts, and
if it is somehow new -- most typically this is taken to mean that emergent
phenomenon displays causal powers not displayed by any of its parts. In
addition to irreducibility, unpredictability, and novelty, it is often asserted
that emergent phenomena are inexplicable -- they defy explanation in terms
of the behaviors of their components. And, as ``explanation" is typically
understood to be explanation by a particular theory, this means that the
behavior of the emergent whole is not fully explained by the theory that
governs the behavior of the component parts.}" (Batterman \cite{Batt} p.1-2)}

{\small  \textit{ ``Almost all conceptions of emergence characterise novelty by appealing to some
conception of predictability or derivability, questioning whether it is possible to
obtain one set of properties from another by some mechanical process of logical
derivation. Kim's 1995 definition is typical from one concerned with the philosophy
of mind: `a property of a complex system is said to be `emergent' just
in case, although it arises out of the properties and relations characterizing simpler
constituents, it is neither predictable from, nor reducible to, these lower-level
characteristics.' The thought is that if it is possible to derive systemic properties
from those of their components, these properties cannot be truly novel -- for they
were "there all along" in the parts, and it was merely a matter of careful analysis
to make this fact manifest. This leads naturally to a distinction between this
`prediction' or `reduction' being one that could be carried out in practice, and one
that is possible only in principle."} (Mainwood \cite{Main} p. 27)} Here lies the distinction between \textbf{ontological} (in principle) vs \textbf{epistemological} (in practice) \textbf{emergence}.


\vskip .2cm
\noindent \textbf{Reductionism- Constructivism}

{\small \textit{ ``The workings of our minds and bodies,
and of all the animate or inanimate matter of which we have any
detailed knowledge, are assumed to be controlled by the same set of
fundamental laws, which except under certain extreme conditions we
feel we know pretty well." }(Anderson \cite{And72} p. 393)}

{\small \textit{`` ... the material composition of organisms is exactly the same as that
found in the inorganic world. Further ... none of the events and processes
encountered in the world of living organism is in any conflict with
the physico-chemical phenomena at the level of atoms and molecules."}
(Mayr's ``constitutive reductionism' in the context of biology \cite{Mayr} p.59 quoted by Mainwood \cite{Main} p. 16}

{\small \textit{ ``... the reductionist hypothesis does not by any means imply a constructionist
one: The ability to reduce everything to simple fundamental
laws does not imply the ability to start from those laws and
reconstruct the universe.}

{\small \textit{
``The constructionist hypothesis breaks down when confronted with the
twin difficulties of scale and complexity.} (Anderson \cite{And72} p. 393)}

\vskip .2cm
\noindent \textbf{Microphysicalism} and the `New Emergentists"

What Anderson and Weinberg refer to as `reductionism' is at odds with the metaphysicians' usage, who call it `microphysicalism'.


{\small \textit{`` In metaphysics, the claim endorsed by Anderson and Mayr is usually called microphysicalism.
[M]icrophysicalism ... is the doctrine that actually (but not necessarily)
everything non-microphysical is composed out of microphysical entities
and is governed by microphysical laws."} (Pettit \cite{Pettit94})}

{\small \textit{`` Pettit also brings out the two components which constitute microphysicalism.
The first is that everything in the empirical world -- every particular
in that world  -- is composed in some sense out of subatomic materials.
And second is that everything that happens in the empirical world
happens, ultimately, under the controlling influence of subatomic forces and laws."} (Pettit \cite{Pettit96} p.342-3)}

Mainwood refers to proponents and adherents of these tenets as the New Emergentists.

On the issue of ``in practice" versus ``in principle"  mentioned above, {\small \textit{``.. the New Emergentists are accused of illegitimately citing evidence for epistemological emergence (a practical difficulty in deriving the properties of a large system) as support for claims of ontological emergence (there exist emergent
properties that are ``novel" in a metaphysically important sense).}  (Mainwood \cite{Main} )}


\vskip .2cm
\noindent \textbf{Supervenience} -- a key concept philosophers use for qualifying emergence.

\textit{\small  ``The most promising approach to capturing the `determination'
inherent in property physicalism is through a relation of supervenience.
Take a set of objects O, let their physical properties form a set $P_A$, and all their
properties (both physical and ostensibly non-physical) form a set $P_B$. To capture
physicalism via supervenience, we demand that the extensions of elements of $P_B$
are fixed by specifying the extensions of the elements of $P_A$. The simplest way of
accommodating any such n-adic relations amongst objects $o_1, o_2, o_3$ is to treat it
as a one-place property of the larger object composed of all of these. So we need
to close the set of objects O under the operation of taking ordered n-tuples.}

{\small \textit{ A little more formally: the properties $P_B$ supervene on $P_A$ with respect to a
set of objects O, on which both are defined, iff any two objects in O that match
with respect to all properties in $P_A$, also match with respect to all properties in
$P_B$. The central claim of property physicalism can then be expressed very simply:
all properties supervene on physical properties."}  (Mainwood  \cite{Main} p. 22)}

Using these technical terms in philosophy, Mainwood summarizes, {\small \textit{
``... systemic properties are novel, if and only if it is practically impossible to
derive them from the microphysical properties mentioned in microphysical supervenience."}
(\cite{Main} p. 30)}

Three proposals of novelty were examined by Mainwood \cite{Main} (Sec. 1.5):
\textit{\small 1) a failure of inter-theoretic reduction; 2) an impossibility of deducing the systemic
properties from the properties of the parts; or 3) a failure of mereological supervenience.}
The approaches appeal to three entirely separate distinctions
between the properties of parts and wholes; thus according to Mainwood, there are three entirely different
sets of criteria for emergence. We leave it for the interested reader to follow his treatise but offer some insights on these issues in a later section.

\subsection{Key words and concepts used by physicists}

\vskip .2cm
\noindent \textbf{Fundamental}:

{\small \textit{ ``The modern theory of critical phenomena has interesting implications for our understanding of what constitutes "fundamental" physics. For many important problems, a fundamental under-
standing of the physics involved does not necessarily lie in the
science of the smallest available time or length scale. The extreme
insensitivity of the hydrodynamics of fluids to the precise physics
at high frequencies and short distances is highlighted when we
remember that the Navier-Stokes equations were derived in the
early nineteenth century, at a time when even the discrete atomistic nature of matter was in doubt.}
( \cite{Nelson Defects} p.3, quoted by Batterman \cite{Batt}]}

\vskip .2cm
\noindent
\textbf{Protectorate:}

According to Laughlin and Pines (LP) \cite{LP}
{\small \textit{.. a (quantum) protectorate, (is) ``a stable state of matter whose generic low-energy properties are determined by a higher organizing principle and nothing else." }

An example of LP's ``higher organizing principle", used also by Anderson earlier,  is spontaneous symmetry breaking. \textit{ \small``.. one does not need to prove the existence of sound in a solid, for it follows from the existence of elastic moduli at long length scales, which in turn follows from the spontaneous breaking of translational and rotation symmetry characteristic of the crystalline state."}  What is more telling is, \textit{ \small ``Conversely, one therefore learns little about the atomic structure of a crystalline solid by measuring its acoustics."}

Morrison \cite{Morr}) 
\textit{\small ``.. referred to them as `theoretical principles' in order to capture the idea that there is a dynamical process associated with these principles responsible for producing certain kinds of behavior."}


In the eyes of the theoretical physicists, `theoretical' is too general a term to identify these principles, and can become nondescriptive.
These principles provide not only how the micro-parts are organized into the meso-`whole', and here we are referring two adjacent rungs in a hierarchy, but also the dynamical process which they come into being.  This would correspond to the specification of -- how to choose or constitute -- the collective variables and how to describe their states and dynamics, the collective behavior, which is the critically important issue.

\vskip .2cm
\noindent  \textbf{Parts and Whole relation}:



I agree with these summary statements by Morrison below and by Batterman next:

\textit{\small ``Not only do we need to reorient our thinking about the role of natural kinds as a method for differentiating essential features of matter, but new ways of thinking about the part/whole relations
involved in defining and describing emergent phenomena are required."}

{\small \textit{ ``And perhaps most importantly, a re-evaluation of reductionist strategies for defining the relationship between different theoretical levels is crucial for making sense of emergence in physics.
In connection with the ¡°levels¡± approach characteristic of this picture
of emergence, it is important to note the differences between emergence
and the fact that phenomena at different scales may obey different fundamental
laws."} (Morrison \cite{Morr} p.886)}

\subsection{Overall Assessment}
\vskip .2cm
{\small \textit{
``A number of philosophers of science have imposed these philosophical conceptions upon physical theory in an attempt to address the issue of emergence in physics. While I believe there are some benefits to this methodological
approach, on the whole I think it is better to turn the process on its head. We
should look to physics and to ``emergent relations" between physical theories
to get a better idea about what the nature of emergence really is. Trying to
impose a conceptual framework designed primarily to deal with the problem
of the mental's relation to the physical is by and large unhelpful." }(Batterman \cite{Batt} p.2)}

I think the same can be said about the relation of science and philosophy of science in general. If one wants to not just talk about the philosophy of a subject X  (the layman's use of the word ``philosophize on" is even worse), but to provide philosophical insights to the key issues of X, in terms of asking better questions and seeing clearer directions, it is imperative that one should first have the technical mastery of X. E.g., one must have lived a life to talk about the philosophy of life.

\section{Understanding emergent phenomena: Some observations and a useful platform}

In this section I wish to make some observations on how philosophers can better communicate with, and of greater help to  physicists, e.g., sorting out the complexity of issues in problems physicist wrestle with and probing deeper into specific key issues in these problems.  Three aspects:

1) \textbf{Conceptual Constructs}: It is not enough to rely on a reductionist conceptual approach, such as what Mainard related to about Russell's impact on philosophers,  that all mathematical statements can be reduced to set theory. Many physical phenomena defy simple descriptions or monolithic construction, some may even appear illogical (in reasoning, not in its intrinsic consistency). A new physical theory will become more logical and its conceptual construct more rational (think special relativity and quantum physics) only after it has been proposed, examined and checked against observations and experiments, not before. Physicists create models to capture the essence of the novel phenomena and they make approximations to test out characteristic behavior in different parameter regimes. What they are wrestling with is physical reality in whatever shape and form it be, raw and often unruly. Physicists approach truth from reality. This process may be aided by abstractions of theories with some sense of completeness and perfection but the final judgments come from experiments. Here we see the difference in attitude stemming from approaching truth in the epistemic rather than the ontological sense. For emergent phenomena, rather than assuming a linear progression of logical deduction or inference we need a conceptual scheme  which can explain the emergence of a hierarchical construct, with a tower of levels of structure, the collective variables most suitable for each level, the laws governing them, and even more demanding, the inter-level meta-constituents and their activities.

2) \textbf{Methodology and Vehicles}: It seems that there is still heavy reliance on the method of close scrutiny of  the internal logical consistency of an argument -- down to every sentence written or word used,  to the extent that a slight detected crack in the logic of an argument immediately spells the demise of the whole theory and the triumph of the protagonist. This golden example set up by the classic philosophers before the advent of modern science is unfortunately dated and inadequate when applied to physical issues, because reality often defies logic, at least in the epistemic level.  This is  in contrast to mathematical constructs,  especially when the challenge is to  come up with a new paradigm to deal with new issues such as emergence.  This methodology is not wrong, it works for disciplines based on pure logic. The mathematical tools  physicists use to derive equations are  constructed with rigorous logic. But there is a divide, often unnoticed or downplayed by non-physicists, between casting physical ideas to mathematical models and inferring physical meanings from mathematical results.  It is just wrongly applied to arguments, and to the extent arguments cannot be represented rigorous by mathematical concepts, this approach can easily lead one to drawing sweeping or overgeneralized statements or/and missing the key issues, defeating the purpose of philosophy.

3) \textbf{Mechanisms and Processes} For the study of emergence in physics, a modest way I can suggest which produces reliable and helpful results for philosophical inquiries is to examine existing clear-cut examples with valid physical theories at both the micro (basic or originating) and macro (effective or emergent) levels, examine the underlying physical issues, extract the commonalities and differences, identify notions, terminology and concepts which need further clarification. Examples are thermodynamics via kinetic theory, hydrodynamics via molecular dynamics, collective excitations in (atomic-based) condensed matter systems; quantum Hall effect \cite{BainQHE},  chiral-dynamics in QCD, dimensional reduction in  Kaluza-Klein theories.  In terms of constructing paradigms for emergence from a thorough understanding of the specifics, in addition to the modeling and approximation schemes physicist use all the time, here, what is particularly relevant are the mechanisms and processes whereby new phenomena emerge at a particular collective (or derived) level from a more basic (or elementary) level. Symmetry breaking mentioned by Anderson has been cited repeatedly by philosophers. There are more, which can be gleaned from the well-established examples mentioned above. We will dwell on a few of these in the following.

A \textbf{useful platform} in physics to explore emergence is nonequilibrium statistical mechanics. It consists  not just of the formalisms or theorems for physical systems out of equilibrium,  not just finding real-time causal equations of motion for the evolution of such systems, but more importantly, for our purpose here, its conceptual constructs in the foundational issues, the physical meaning they convey, but also the  methodology adapted to specific physical setups, and the specific mechanisms and processes. They all prove to be essential for the explanation of the emergence of forms and structures, even new symmetry principles and constituting laws. Just as equilibrium statistical mechanics, in particular Gibbs' ensemble theory,  has acted as the base for condensed matter physics for almost a century, nonequilibrium statistical mechanics  is the underpinning of disciplines from soft matter physics to chemical-biological systems to sociological- collective behavior to financial-economic modeling. It is in these complex systems where emergent phenomena manifest most vividly and strikingly.

The key issues of nonequilibrium statistical mechanics, which includes stochastic mechanics, open system dynamics, etc, are:   (i) collectivity, coarse-graining, correlations - coherence;   (ii) noise, fluctuations, stochasticity;  (iii) nonlinearity, nonMarkovianity, nonlocality. They all have a bearing on emergence in varying degrees.  We can discuss only one issue here from this list, that of coarse-graining and persistent structures (``protectorates"). A description of various physical systems  with reference to how certain information in the system is kept, neglected, lost or degraded is facilitated by the distinction between closed, open and effectively open systems. This aspect which underlies the thematic material of the next section has been discussed before (see, e.g, the last sections in \cite{timeasy}, excerpts appeared in Chapter 1 of \cite{CH08} with some reference updating). We learn that even the very first step in defining an open system or an effectively open but otherwise closed system, is non-trivial. In fact the judicious choice of an appropriate set of collective variables already provides half of the answer to the issues of emergence. (E.g., witness how simple yet powerful the Laughlin wave function is to capture the fractional quantum Hall effect.) The remaining half comes from deriving the dynamics of these collective variables in describing the physical phenomena at this level of structure and interactions (e.g., thermodynamics in terms of temperature, chemical potential, entropy, enthalpy).

\section{Key physical issues worthy of closer attention by philosophers: 3Cs and 3Ds}

The 3Cs are ``Collectivity, Coarse-graining, Correlation- Coherence". The 3Ds are: Details, Details, Details \footnote{I'm not referring to what is contained in the title of \cite{BattBook}, but to the details in a physical phenomenon and in a physical theory attempting to explain such phenomena.}: minute details in the study of emergence could be important. We feel that in a new field which lacks a well-defined paradigm it is important to pay attention to all the details before drawing some general conclusion than concocting general principles from pure reasons in the belief that they can explain or even predict all the details \footnote{A section meant to illustrate this point with the processes of environment-induced quantum decoherence (See, e.g., \cite{PazZurLH,ZurRMP}) in the emergence of the classical world is omitted here due to space limitation. What I wanted to demonstrate with this example is, even when a process or mechanism can be and has been identified, different parameter regimes of the factors involved (e.g., low temperature, super-Ohmic spectral density) can yield very different outcomes, see, e.g., \cite{PHZ93}. In the decoherent histories approach the existence of quasi-classical domains \cite{GelHar93} is, contrary to what was initially conjured, a very complex issue. See, e.g, \cite{DowKent}.}.

We shall focus on the issue of coarse-graining and persistent structures here. There is no space to include the second set of issues, namely, correlation, and for quantum systems, quantum coherence and entanglement. These are important issues for understanding  how one level of structure connects to, or yields / reduces to another. They appear especially acutely for strongly correlated systems or systems with memories because they make the choice of coarse-graining measures particularly difficult \footnote{From the projection operator \cite{Zwa01} viewpoint, this corresponds to situations where, confronted with the integro-differential equation for one system (containing the dynamics of the other subsystem), one can find no easily justifiable or implementable approximations.}.  Some discussions of this point can be found in \cite{MQP2} which addresses these aspects of macroscopic quantum phenomena with references.

\subsection{Elimination of degrees of freedom (DOF) and emergence in effective field theory (EFT)}

To set the stage for our discussions I quote the Abstract of Bain \cite{BainEFT} who used two examples in physics \cite{BainQHE}  to illustrate some issues of emergence in effective field theory (EFT) (see, e.g., \cite{PolEFT}).  As stated in the beginning, the more specific a physical theory or model a philosopher can relate to, the more helpful it is to physicists who want to see how related philosophical issues are being dealt with. The issue of the `elimination of degrees of freedom' raised there by Bain and by Wilson \cite{Wilson}, as related by Bain, are relevant to our theses discussed in the next subsection, namely, on the robustness of coarse-graining measures and persistent structures.

Bain \cite{BainEFT} suggests that \textit{\small EFTs satisfy the following disiderata for a notion of emergence:}

\textit{\small (i) Emergence should involve microphysicalism, in the sense that the emergent system
should ultimately be composed of microphysical systems that comprise the
fundamental system and that obey the fundamental system's laws. (ii) Emergence should involve novelty, in the sense that the properties of the emergent system should not be deducible from the properties of the fundamental system.}

\textit{\small These disiderata are underwritten in an EFT by the elimination of degrees of freedom in
its construction. Thus the properties of a system described by an effective Langrangian
density Leff can be said to emerge from a fundamental system described by a high energy
Langrangian density L in the following sense:
(a) High-energy degrees of freedom are integrated out of L. This secures
microphysicalism insofar as it entails that the degrees of freedom of $L_{eff}$ are exactly
the low-energy degrees of freedom of L.
(b) $L_{eff}$ is expanded in a local operator expansion (either to guarantee locality or as a
means of approximating the functional integral in (a)). The result is dynamically
distinct from L, and this secures novelty in the sense of the failure of lawlike
deducibility from L of the properties described by $L_{eff}$.}

\textit{\small According to Mainwood (\cite{Main} pp. 107, 116), the mechanisms Anderson and Laughlin and Pines identify as underwriting (i) and (ii) are spontaneous symmetry breaking and universality .
Neither of these is generally applicable to EFTs, as the Quantum Hall liquid example indicates.}

\textbf{Elimination of DOF} \textit{\small Mainwood (\cite{Main} p. 284) further suggests that a nontrivial notion of
emergence requires the specification of a physical mechanism to underwrite (i) and (ii),
and it might seem that the elimination of degrees of freedom in an EFT is a formal, as
opposed to a physical, mechanism.}

As related to by Bain, \textit{\small Wilson \cite{Wilson} similarly identifies the elimination of degrees of freedom (DOF) as an essential characteristic of a notion of emergence. For Wilson, DOF elimination plays
two roles. First it secures the physical acceptability of an emergent entity by securing
the lawlike deducibility of the entity's behavior from its composing parts (p. 295), and such physical acceptability partially underwrites physicalism. Second, according to Wilson, DOF elimination entails that an emergent entity is characterized by different law-governed properties and behavior than those of its composing parts,
and this suggests that the former cannot be reduced to the latter (p. 301). This failure of ontological reduction might charitably be associated with a notion of novelty (although Wilson's explicit goal is simply to establish peaceful coexistence between physicalism and non-reductivism).}

\textbf{Bain's comment:}
\textit{\small ``This might suggest similarity with the above account of emergence in EFTs. However, the type of DOF elimination involved in the construction of an EFT is distinct from Wilson's notion in two major respects. First,
DOF elimination in an EFT is typically characterized by a failure of lawlike
deducibility: The lawlike behavior of entities described by an EFT cannot, in general,
be deduced from the lawlike behavior of the entities described by its high-energy theory.
This failure, I suggested above, is what underwrites a notion of novelty. Second, in the
presence of such failure, physicalism is preserved, insofar as, in DOF elimination in an
EFT, the degrees of freedom of the EFT are exactly the low-energy degrees of freedom
of the high-energy theory."}

\textbf{Our comments:}  Emergence of a new (macro) level of structure does not obtain simply by the ``elimination" of (micro) degrees of freedom. This emerged level of structure usually comes from a new set of collective variables which bears no relation to the variables of the sub-level. An easy example is again thermodynamics, where one important collective variable is the temperature T. The DOF of the micro theory are the coordinates and momenta $(q_i, p_i)$ of N ($N_A$ Avagadro's number $10^{23}$) molecules. Elimination of this huge number of DOF to a few does not lead one any closer to seeing the foremost collective variable T. It is in fact the wrong way to think and to go.   Temperature  is related to the kinetic energy of the  molecules, not extractable from the elimination of these variables.  The judicious  choice, or more often than not, the creative construction of a set of collective variables  from a given set of micro-variables is required and usually poses the most critical challenge.

The following two subsections excerpted from \cite{timeasy} reflect the complexity of this issue.

\subsection{Measures of Coarse-graining}

Coarse graining in the most general sense refers to some information
lost, removed, or degraded from a system. It could come about because
these information is \textit{inaccessible to us}, due to the limited
accuracy in our observation or measurement. A drastic example is
Planck scale physics, the details of which are mostly lost (hard to
retrieve) because the world we live in today is an ultra-low energy
construct. For this one needs to invoke ideas like effective field
theory \cite{PolEFT}. Even when information is fully accessible
to us in principle, in practice one may only be interested in some
aspects of the system.  \textit{We choose to ignore} certain
variables such as ignoring the higher order correlations in
Boltzmann's kinetic theory, or ignoring the phase information in a
quantum system by imposing a random phase approximation. We do this
by `integrating over' or ``projecting out' these `irrelevant'
variables.


Here are some examples of coarse-graining in action. We start with
the familiar Boltzmann theory: implementation of the molecular chaos
assumption (i.e., the 2 particle distribution function $f_2 = f_1
f_1$ can be expressed schematically as a product of two 1-particle
distribution functions $f_1$ ) entails performing a coarse-graining
in the collision integral of space over the range of interaction and
of time over the duration of a collision.



Note also that coarse-graining is a necessary but not sufficient
condition for entropy generation. It does not always produce a
dissipative system. Truncation of the BBGKY hierarchy leads to a
closed subsystem composed of $n$- particle correlation functions
whose dynamical equations are unitary. (An example mentioned before
is the Vlasov equation describing particle interaction via long range
forces.) In quantum field theory equations derived from a finite-loop
effective action are also unitary -- at one loop the effect of the
quantum field  on the particles manifests through the renormalize
masses and charges  (to be exact, the equations of motion
derived from a finite-loop effective action are unitary if none of
the relevant correlation functions are `slaved' \footnote{See \cite{CH08} Chapter 6,
Sec. 3 and Chapter 9 Sec. 2.3 for a discussion of this concept.} for
$\ell$ loops, one must keep the first $\ell +1 $th order
correlations, otherwise dissipation in the sense defined above sets
in  - dissipation  is absent only in very specific situations, such as
a free theory or equilibrium initial conditions). That is perhaps
why (if one limits one's attention to loop expansions) statistical
mechanical concepts rarely came to the fore, until one starts asking
questions of a distinct nature, such as how dissipative dynamics
appears in an otherwise unitary system, and the origin and nature of
noise in quantum field theory. A causal condition need be introduced
to render the dynamics of the subsystem irreversible. This opens up
another important theme: effective field theory viewed
in the open system framework (See \cite{CHEFT} for an example of how noise from the higher energy sector
can serve as a measure of the validity of a low energy effective field theory.)

\subsection{Persistent structures in the physical world -- `protectorates'}

We have seen from the above discussions that the appearance of
irreversibility is often traced to the initial condition being
special in some sense. The dynamics of the system and how it
interacts with its environment also enter in determining whether the
system exhibits mixing or dissipative behavior. For the sake of
highlighting the contrast we could broadly divide the processes into
two classes depending on how sensitive they are to the initial
conditions versus the dynamics.  One can say that the first class is {\it a priori}
determined by the initial conditions, the other is {\it a posteriori}
rather insensitive to the initial conditions. Of the examples we have
seen, the first group includes divergent trajectories in molecular
(micro) dynamics, Landau damping, vacuum particle creation, the
second class includes gas (macro) or fluid dynamics, diffusion, particle creation with
interaction, decoherence. Appearance of dissipation is accompanied by
a degradation of information via coarse graining (such as the
molecular chaos assumption in kinetic theory, restriction to
one-particle distribution in particle creation with interaction,
`integrating out' some class of histories in decoherence).


Many perceived phenomena in the observable physical world, including the phenomenon of time-asymmetry, can be understood in the open-system viewpoint via the approximations introduced to the objective microscopic world by a macroscopic observer.\cite{QdecBook,OmnesBook,Peres}

We have discussed the procedures such as the system-environment split and the
coarse-graining of the environment which can bring about these
results. However, a set of more important and challenging issues
remain largely unexplored, i.e.,  under what conditions the
outcomes become less subjective and less sensitive to these
procedures. These procedures provide one
with a viable prescription to get certain general qualitative
results, but are still not specific and robust enough to explain
how and why the variety of observed phenomena in the physical
world arise and stay in their particular ways. To address these
issues one should ask a different set of questions:\\

{\it 1) By what criteria are the system variables chosen?
-- collectivity and hierarchy of structure and interactions}

\vskip .2cm

In a model problem, one picks out the system variables -- be it
the Brownian particle or the minisuperspace variables -- by fiat.
One defines one's system in a particular way because one wants to
calculate the properties of that particular system.  But in the
real world, certain variables distinguish themselves from others
because they possess a relatively well-defined, stable, and
meaningful set of properties for which the observer can carry out
measurements and derive meaningful results. Its meaningfulness is
defined by the range of validity or degree of precision or the
level of relevance to what the observer chooses to extract
information from. In this sense, it clearly carries a certain
degree of subjectivity-- not in the sense of arbitrariness in the
exercise of free will of the observer, but in the specification of
the parameters of observation and measurement. For example, a
thermodynamic  variable like temperature provide excellent description of systems  close to equilibrium; in other regimes one needs to describe the system in terms of kinetic-theoretical or molecular-dynamical variables where temperature carries no direct meaning.


The level of relevance which defines one's system changes with the level of structure of matter and the relative importance of the forces at work at that level. The improvement of the Weinberg-Salam model with $W, Z$ intermediate bosons over the Fermi model of four-point interactions is what is
needed in probing a deeper level of interaction and structure
which puts the electromagnetic and weak forces on the same
footing. Therefore, one needs to explore the rules for the
formation of such relatively distinct and stable levels, before
one can sensibly define one's system (and the environment) to
carry out meaningful inquiries of a statistical nature.

What is interesting here is that these levels of structures and
interactions come in approximate hierarchical order (thus, e.g., one doesn't
need QCD to calculate the rotational spectrum of a nucleus, and the
manifold picture of spacetime will hopefully provide most of
what we need in the post-Planckian era). One needs both some
knowledge of the hierarchy of interactions and the way effective theories emerge from
`integrating out' variables at very different energy scales in the
hierarchical structure (e.g., ordinary gravity plus particle theory
regarded as a low energy effective higher-dimension or Kaluza-Klein
theory) The first part involves fundamental constituents and
interactions and the second part the application of statistical
methods. One should also keep in mind that what is viewed as
fundamental at one level can be a composite or statistical mixture
at a finer level. There are system-environment separation schemes
which are designed to accommodate or reflect these more intricate
structures, from the mean field-fluctuation field split to the
multiple source or nPI formalism (see \cite{CH08} Chapter 6)  for the
description of the dynamics of correlations and fluctuations. The validity of these
approximations depends quite sensitively on where exactly one wants
to probe in between any two levels of structure. Statistical
properties of the system such as the appearance of dissipative
effects and the associated irreversibility character of
the dynamics in an open system certainly depend on this separation.\\

{\it 2)  How does the behavior of the subsystem depend on
coarse-graining? -- sensitivity and variability of
coarse-graining, stability and robustness of emergent structure}

\vskip .2cm

Does there exist a common asymptotic regime as the result of
including successively higher order iterations in the same
coarse-graining routine? This measures the sensitivity of the end
result to a particular kind of coarse-graining. How well can
different kinds of coarse-graining measure produce and preserve the
same result? This is measured by its variability. Based on these
properties of coarse-graining, one can discuss the relative
stability of the behavior of the resultant open system after a
sequence of coarse-grainings within the same routine, and its
robustness with respect to changes to slightly different
coarse-graining routines.

    Let us illustrate this problem with some simple examples.
When we present a microscopic derivation of the transport
coefficients (viscosity, heat conductivity, etc) in kinetic theory
via the system-environment separation scheme, we usually get the
same correct answer independent of the way the environment is chosen
or coarse-grained. Why? It turns out that this is the case only if
we operate in the linear-response regime. (See
\cite{FEVE63}). The linear coupling between the system and the
environment makes this dependence simple. This is something we
usually take for granted, but has some deeper meaning. For nonlinear
coupling, the above problem becomes nontrivial. Another aspect of
this problem can be brought out \cite{BalVen,Spohn} by comparing
these two levels of structure and interaction, e.g., the
hydrodynamic regime and the kinetic regime. Construct the relevant
entropy from the one-particle classical distribution function $f_1$,
that gives us the kinetic theory entropy $S_{kt}$ which is simply
$-k H_{B}$, where $H_{B}$ is the Boltzmann's H-function. Now compare it with
the hydrodynamic entropy function $S_{hd}$ given in terms of the
hydrodynamic variables (in this case, the number and energy
density), one sees that $S_{hd} > S_{kt}$. A simple physical
argument for this result is that the information contained in the
correlations amongst the particles are not included in the
hydrodynamic approximation. Even within the kinetic theory regime
there exist intermediate stages described by suitably chosen
variables \cite{Spohn}. The entropy functions constructed therefrom
will reflect how much fine-grained information is lost. In this
sense $S_{hd}$ is a maximum in the sequence of different
coarse-graining procedures. In the terminology we introduced above,
by comparison with the other regimes, the hydrodynamic regime is
more robust in its structure and interactions with respect to
varying levels of coarse-graining.  One way to account for this is,
as we know, the hydrodynamic variables enter in the description of
systems in equilibrium and they obey conservation laws
\cite{HLM95,Bru96,Hal98}. Further coarse-grainings on these systems is expected
to produce the same results, i.e., the hydrodynamic regime is a
limit point of sorts after the action from a sequence of
coarse-grainings. Therefore, a kind of `maximal entropy principle'
with respect to variability of coarse-graining is one way where
thermodynamically robust systems can be located.

While including successively  higher orders of the same
coarse-graining measure usually gives rise to quantitative
differences (if there is a convergent result, that is, but this
condition is not guaranteed, especially if a phase transition
intervenes), coarse-graining of a different nature will in
general result in very different behavior in the dynamics of the
open system. Let us look further at the relation of variability
of coarse-graining and robustness of structure.

Sometimes the stability of a system with respect to the variability
of coarse-graining is an implicit criterion behind the proper
identification of a system. For example, Boltzmann's equation
governing the one-particle distribution function which gives a very
adequate depiction of the physical world is, as we have seen, only
the lowest order equation in an infinite (BBGKY) hierarchy. If
coarse-graining is by the order of the hierarchy -- e.g., if the
second and higher order correlations are ignored, then one can
calculate without ambiguity the error introduced by such a
truncation. The dynamics of the open system which includes
dissipation effects and irreversible behavior will not change
drastically if one uses a different (say more fine-grained)
procedure, such as retaining the fourth order correlations (if the
series converges, which is a non-trivial issue (see, e.g.,
\cite{Dorfman81}). Consider now a different approximation: For a
binary gas of large mass discrepancy, if one considers the system as
the heavy mass particles, ignore their mutual interactions and
coarse-grain the effect of the light molecules on the heavy ones,
the system now behaves like a Brownian particle motion described by
a Fokker-Planck equation. We get a qualitatively very different
result in the behavior of the system.

In general the variability of different coarse-grainings in
producing a qualitatively similar result is higher (more
variations allowed) when the system one works with is closer to a
stable level in the interaction range or in the hierarchical
order of structure of matter. The result is more sensitive to
different coarse-graining measures if it is far away from a
stable structure, usually falling in between two stable levels.
Only robust systems survive in nature and carry definite meaning in terms of their persistent structure and systematic evolution. This is where the relation of coarse-graining and persistent structures enters.


So far we have only discussed the activity around one level of robust
structure. To investigate the domain lying in-between two levels of
structures (e.g., between nucleons and quark-gluons) one needs to
first know the basic constituents and interactions of the two levels.
This brings back our consideration of levels of structures above.
Studies in the properties of coarse-graining can provide a useful
guide to venture into the often nebulous and elusive area between the
two levels and extract meaningful results pertaining to the
collective behavior of the underlying structure. But one probably
cannot gain new information about the fine structure and the new
interactions from the old just by these statistical measures. (cf.
the old bootstrapping idea in particle physics versus the quark
model).







\ack

{\small I thank Gerhard Grossing for his invitation and hospitality.  I am grateful to Professor Hua-Tung Nieh,  Director of the Institute for Advanced Study at Tsing Hua University, Beijing, China for making the arrangements for  my visit in April-May 2011, Professor Zheng-Yu Weng of IAS-THU for showing me how in his high-$T_c$ superconductivity theory emergence of collective variables and their dynamics comes about and Professor Lu Yu of the Institute of Physics, Chinese Academy of Natural Sciences for a discussion on the proper Chinese technical terminology for emergence. This kind of non mission-driven, non utilitarian work addressing purely intellectual issues is not expected to be supported by any U.S. grant agency.}  \\


\end{document}